# The Influence of Fin Shape on the Amplitude of Random Telegraph Noise in the Subthreshold Regime of a Junctionless FinFET

**M. M. Khalilloev[a]*, B. O. Jabbarova[a], and A. A. Nasirov[b]**

[a] *Urgench State University, Urgench, 220100 Uzbekistan*
[b] *National University of Uzbekistan, Tashkent, 100200 Uzbekistan*
**e-mail: x-mahkam@mail.ru*

**Abstract**—The dependence of random telegraph noise (RTN) amplitude on the gate overdrive in a junctionless field-effect transistor (FinFET) with rectangular and trapezoidal channel (fin) cross sections manufactured using silicon-on-insulator technology has been numerically simulated. It is established that the RTN amplitude in the subthreshold region of gate voltages for a FinFET with a trapezoidal cross section of channel is significantly lower than that for the transistor with rectangular cross section of a channel. In addition, under the same conditions, the RTN amplitude at the threshold gate voltage in a junctionless FinFET is significantly lower than that in planar fully depleted and in usual FinFET.

An effective method to achieve ultralow power consumption consists in using nanosized metal-oxide-semiconductor field effect transistor (MOSFET) in subthreshold logic [1]. In this case, low power consumption is achieved due to MOSFET transistors operating in the sub- or near-threshold regime with currents on a nanoampere level. However, scaling of MOSFET is accompanied by various degradation phenomena. These include short-channel effects [2] and increased sensitivity of transistor characteristics to local charge embedded in the dielectric layer [3, 4] or single charge in the gate dielectric or at its interface with the transistor channel, which gives rise to random telegraph noise (RTN) in the drain current [5]. To suppress these short-channel effects in MOSFET with dimensions reduced to the nanoscale level, special three-gate fin field-effect transistors (FinFETs) were developed; more recently, alternative devices with channel lengths below 20 nm were proposed as junctionless (JL) FinFETs [2]. The technology of JL FinFETs is much simpler and cheaper since it excludes the formation of drain and source regions and eliminates the corresponding problems related to doping of these nanosized regions. The JL FinFET comprises a heavily doped (typically, within $10^{18}$–$10^{19}$ cm$^{-3}$) homogeneous thin semiconductor resistor with insulated gate that controls the flow of charge carriers in this resistor between the source and drain. In contrast to the conventional MOSFET operating in the inversion regime, the JL FinFETs operate in the accumulation mode.

While the short-channel effects in JL FinFETs have been extensively investigated [6–8], issues related to the influence of single oxide- or interface-trapped charges (inducing RTN signals in these transistors) on the drain current and to the application of JL FinFETs in ultralow power logic are studied to a much lower extent. In some works, the main attention of researchers was devoted to dependence of the RTN amplitude on the transistor manufacturing technology, temperature, the gate and drain voltages in the accumulation regime [9, 10] and to dependence of the efficiency of JL FinFET operation in ultralow power logic on the level of channel doping, dielectric constant of spacer material, and channel resistance [11]. Only a few works have reported on the influence of technological factors such as fluctuations of geometric parameters and deviations from preset shapes of nanosized JL FinFET on their noise characteristics. However, these fluctuations can lead to significant variation of the transistor parameters and characteristics. In particular, it was established [12–14] that variations of the characteristics, parameters, and short-channel effects depend on the cannel (fin) shape and dimensions of FinFETs operating in the inversion mode.

In the present work, we have numerically calculated the amplitude of RTN induced by a single charge trapped on a boundary defect at the fin top center of JL FinFETs with channels of different shapes. Figure 1 shows a schematic diagram of this model transistor as manufactured using silicon-on-insulator (SOI) technology with channels in the form of truncated triangular prism and parallelepiped, having rectangular and trapezoidal cross sections, respectively (Figs. 1a, 1b).

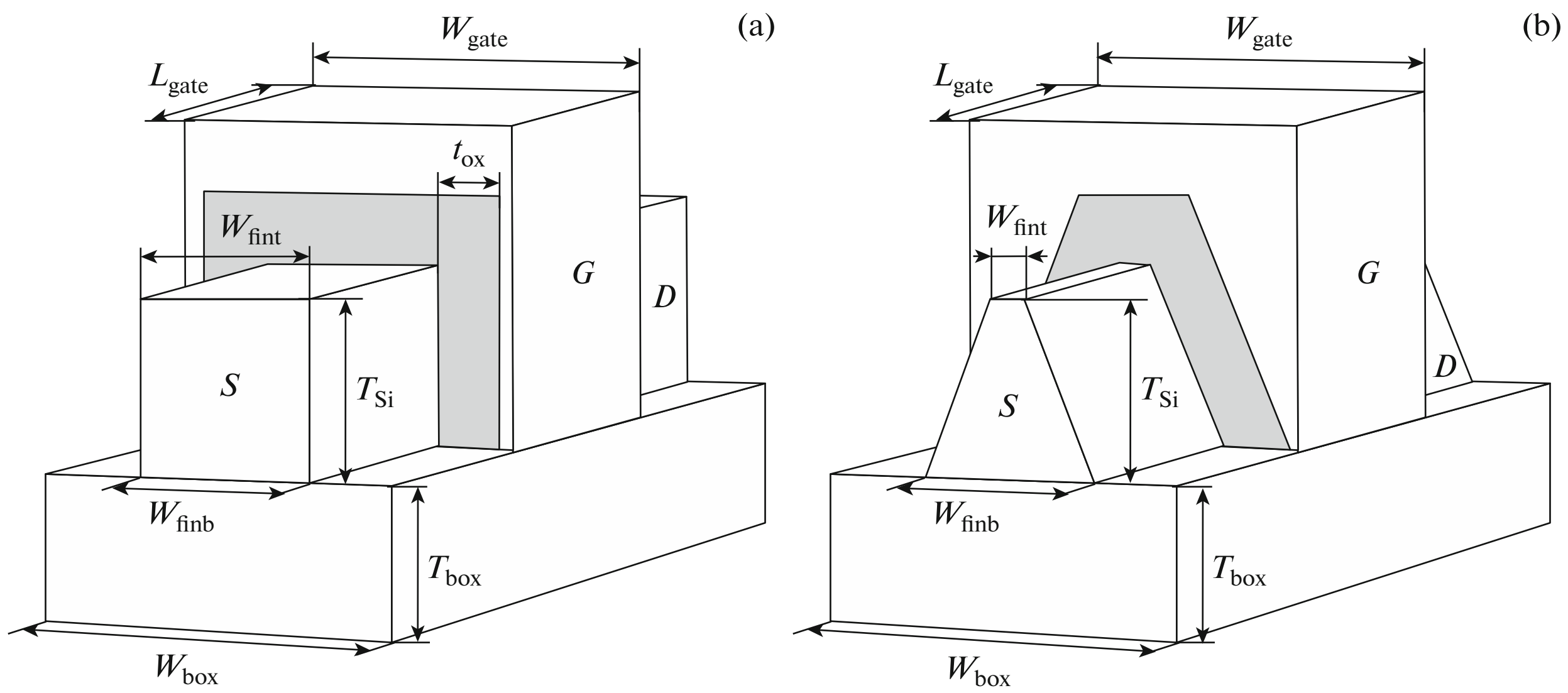

**Fig. 1.** Model structures of JL FinFET with (a) rectangular and (b) trapezoidal fin cross sections (see the text for notation).

Three-dimensional (3D) simulations were performed in the framework of a drift-diffusion model using the Advanced TCAD Sentaurus program package [15]. The model took into account the dependence of carrier mobility on the doping level, the saturation of carrier velocity, and the influence of normal field component on the drain current. Since the transistor had nanoscale dimensions, it was also necessary to take into account the quantum confinement effects.

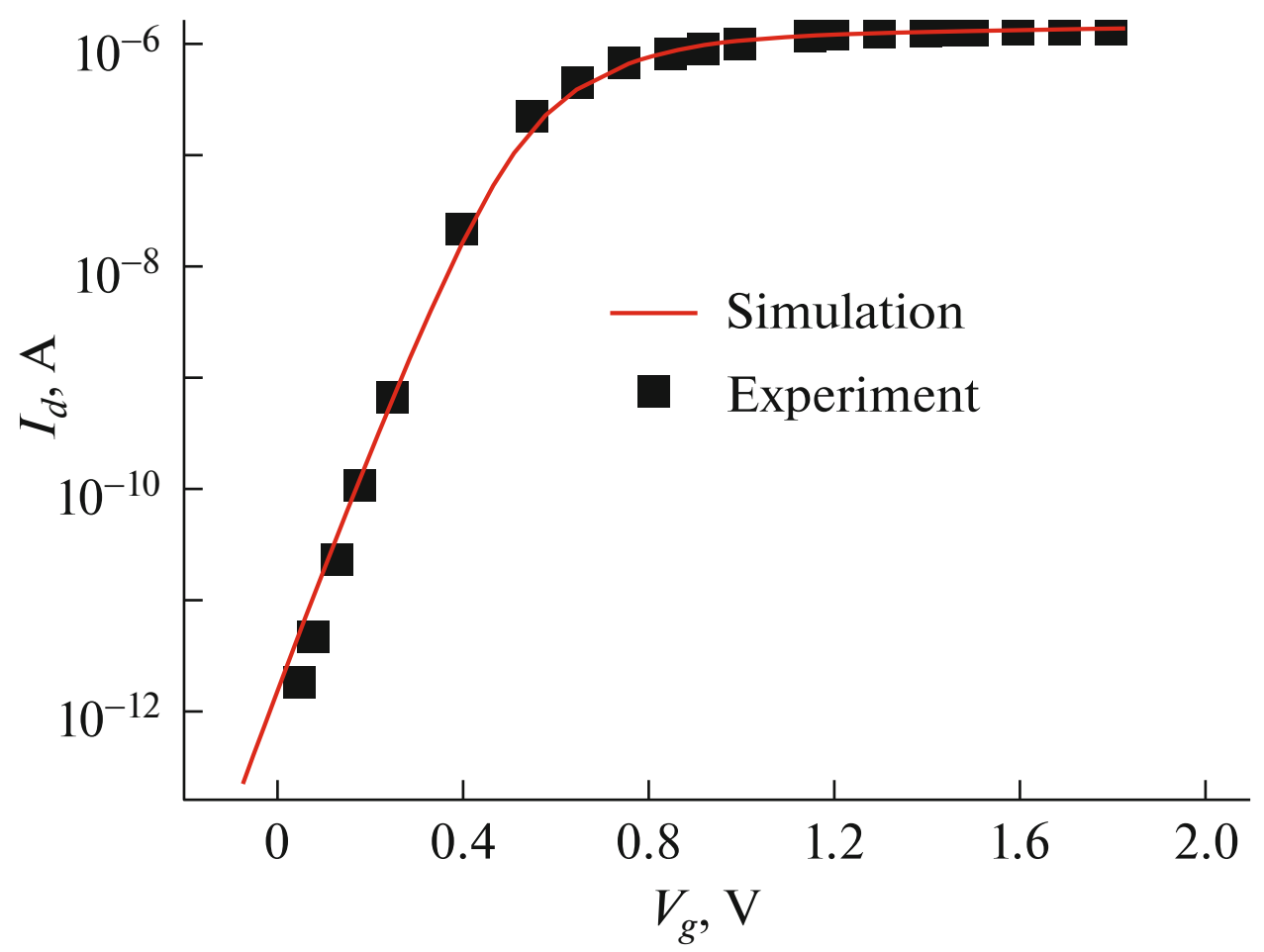

**Fig. 2.** Plot of drain current $I_d$ vs. gate voltage $V_g$ for simulated transistor (solid line) and experimental sample [18] (points) with the same dimensions and parameters.

The present work employed quantum corrections with respect to the density gradient, which are preferred in the framework of drift-diffusion simulations [16, 17]. The adopted model was calibrated according to the experimental data of Barraud et al. [18] (Fig. 2). Numerical calculations were performed for the following parameters. TiN gate ($G$) length $L_{gate}$ = 30 nm (see Fig. 1). The gate insulator was a hafnium oxide ($HfO_2$) layer with equivalent thickness $t_{eff}$ = 0.9 nm. The buried oxide layer had width $W_{box}$ = 48 nm and thickness $T_{box}$ = 25 nm. The fin shape was characterized by top width $W_{fint}$, which was 5 nm for the truncated triangular prism. The fin was doped with boron to a concentration of $5 \times 10^{18}$ cm$^{-3}$ and had a base width of $W_{finb}$ = 10 nm.

Figure 3 shows the results of simulations of the dependence of RTN amplitude on gate voltage $V_g$ excess over threshold voltage $V_{th}$ (i.e., on the difference $V_g - V_{th}$) in a JL FinFET with rectangular and trapezoidal fin cross sections. As can be seen from these data, the RTN amplitude at the gate voltage equal to or less than the threshold voltage is significantly greater than the RTN amplitude at gate voltages above the threshold. In addition, the maximum noise amplitude in the transistor with rectangular channel is almost twice as large as that in the case of a trapezoidal channel. The latter is due to the fact that the distance from single interface-trapped charge occurring in the middle of the fin top to the section with maximum drain density for the channel with rectangular cross section (Fig. 3a) is smaller than that for the trapezoidal cross section (Fig. 3b). At the same time, the observed RTN amplitudes in the near-threshold regime ($V_g = V_{th}$) for JL FinFETs with both fin shapes are significantly lower than the noise amplitude for planar

**Table 1.** Maximum RTN amplitude in MOSFETs manufactured using various technologies

| No. | Technology | Geometric parameters (Fig. 1) | | | | | | Reference | RTN maximum amplitude ($V_g = V_{th}$), % |
|---|---|---|---|---|---|---|---|---|---|
| | | $L_g$, nm | $t_{eff}$, nm | $T_{box}$, nm | $W_{box}$, nm | $T_{Si}$, nm | $W_{Si}$, nm | | |
| 1 | Bulk MOSFET | 22 | 0.9 | | | | 22 | [19] | 28 |
| 2 | FDSOI | 22 | 0.9 | 10 | 10 | 6 | 10 | [19] | 26 |
| 3 | FinFET | 22 | 0.9 | 25 | 10 | 20 | <10 nm | [19] | 8–22 |
| 4 | JL FinFET | 30 | 0.9 | 25 | 48 | 20 | $W_{fint} = 10$ | This work | 2 |
| | | | | | | | $W_{fint} = 5$ | This work | 1.5 |

MOSFET on bulk crystals, a transistor with fully depleted channel (FDSOI), and an SOI FinFET with approximately the same parameters [19] (see Table 1). Even in the subthreshold regime ($V_g < V_{th}$), the RTN amplitude in SOI-JL FinFETs is lower than that in the aforementioned devices in the threshold regime ($V_g = V_{th}$). Table 1 presents the noise amplitude values for all these cases with RTN induced by single interface trapped charge localized at the fin top center.

In concluding, the results of numerical simulations show that the RTN amplitude in SOI-JL FinFETs depends on the fin shape, being significantly lower in the transistor with trapezoidal channel cross section than that in the transistor with rectangular channel cross section. At a threshold gate voltage, the RTN amplitude in SOI-JL FinFETs is lower than that in planar MOS, FDSOI, and junction FinFETs with approximately identical dimensions. This indicates the advantages of using SOI-JL FinFETs with a trapezoidal channel cross section in subthreshold logic with ultralow power.

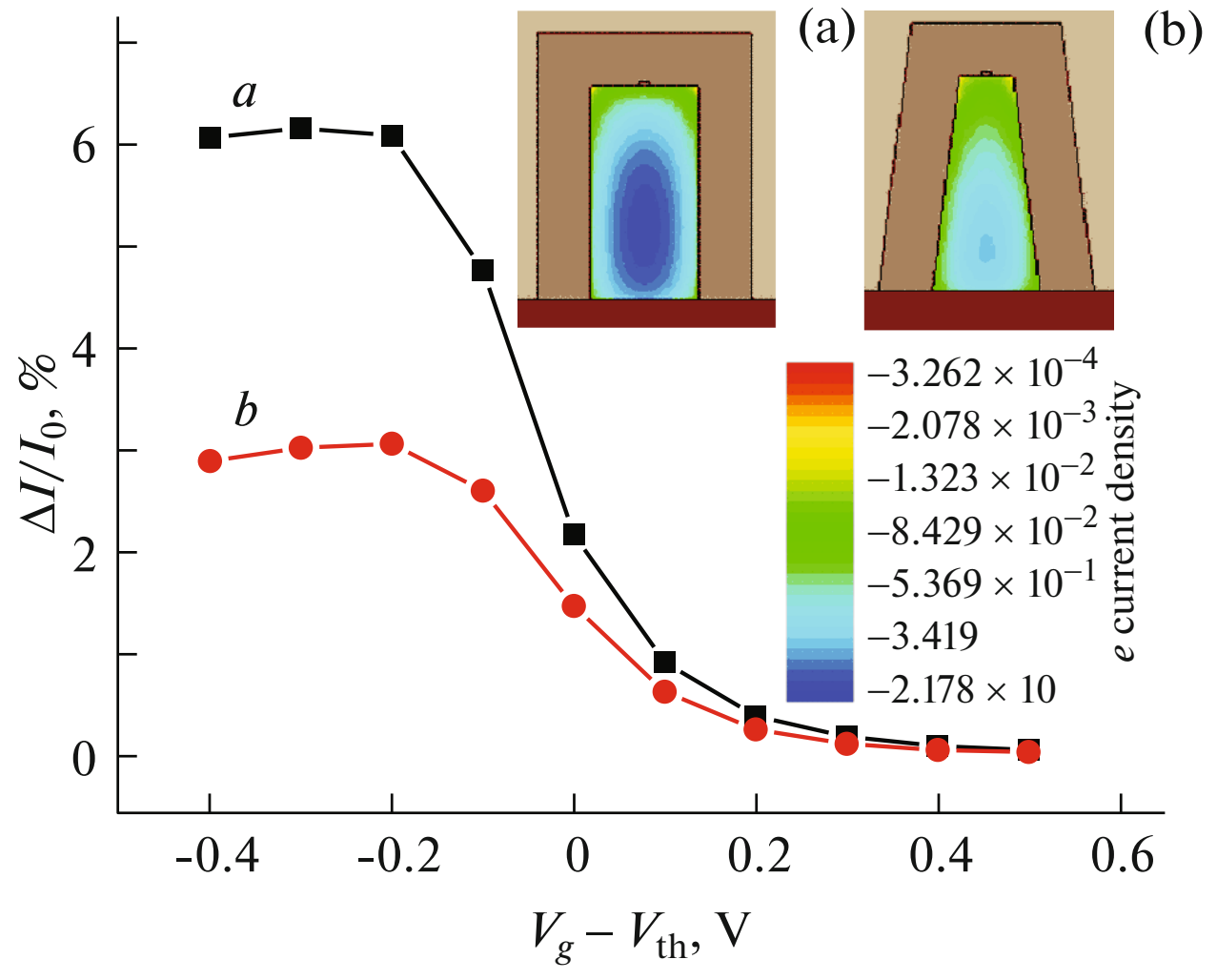

**Fig. 3.** Dependences of the RTN amplitude on the gate overdrive in model junctionless FinFETs with (a) rectangular and (b) trapezoidal fin cross sections. Insets (a) and (b) show the corresponding spatial distributions of electron current density at the center of the fin cross section.

## CONFLICT OF INTEREST

The authors declare that they have no conflict of interest.

*Translated by P. Pozdeev*